\documentclass{emulateapj-rtx4}
\usepackage{amssymb, amsmath, epsfig}

\newcommand{\cMpc}{{\rm cMpc}}

\newcommand{\HI}{H{\sc \, i}}
\newcommand{\HII}{H{\sc \, ii}}
\newcommand{\HeI}{He{\sc \, i}}
\newcommand{\HeII}{He{\sc \, ii}}
\newcommand{\HeIII}{He{\sc \, iii}}

\newcommand{\OVI}{O{\sc \, vi}}
\newcommand{\OVII}{O{\sc \, vii}}

\newcommand{\Myr}{{\rm Myr}}

\begin{document}

\title{Constraints on X-ray Emissions from the Reionization Era}
\author{Matthew McQuinn\altaffilmark{1}}

\altaffiltext{1} {Department of Astronomy, University of California, Berkeley, CA 94720, USA; mmcquinn@berkeley.edu\\}
      
\begin{abstract}         
We examine the constraints on soft X-ray photon emissions from the reionization era.  It is generally assumed that the Universe was reionized by ultraviolet photons radiated from massive stars.  However, it has been argued that X-ray photons associated with the death of these stars would have contributed $\sim 10\%$ to the total number of ionizations via several channels.  The parameter space for a significant component of cosmological reionization to be sourced by X-rays is limited by a few observations.  We revisit the unresolved soft X-ray background constraint on high-redshift X-ray production and show that soft X-ray background measurements significantly limit the contribution to reionization from several potential sources:  X-rays from X-ray binaries, from Compton scattering off supernovae-accelerated electrons, and from the annihilation of dark matter particles.  We discuss the additional limits on high-redshift X-ray photon production from (1) $z\sim3$ measurements of metal absorption lines in quasar spectra, (2) the consensus that helium reionization was ending at $z\approx 3$, and (3) measurements of the intergalactic medium's thermal history.  We show that observations of $z\sim3$ metal lines allow little room for extra coeval soft X-ray emission from a nonstandard X-ray sources.  In addition, we show that the late reionization of helium makes it quite difficult to also ionize the hydrogen at $z>6$ with a single source population (such as quasars) and that it likely requires the spectrum of ionizing emissions to soften with increasing redshift.  However, we find that it is difficult to constrain an X-ray contribution to reionization from the intergalactic temperature history.  We show that the intergalactic gas would have been heated to a narrower range of temperatures than is typically assumed at reionization, $2-3 \times 10^4~$K, with this temperature depending weakly on the ionizing sources' spectra.  
\end{abstract}


\section{Introduction}

Extreme ultraviolet (EUV) radiation from stars within diminutive galaxies is the leading candidate for what sourced hydrogen reionization at $z\sim 10$ (e.g., \citealt{wyithe03}).  This scenario has been the focus of the majority of theoretical and numerical work on reionization (e.g., \citealt{furlanetto04, iliev06, mcquinn07, trac07, finlator11b}).  However, other models may still be viable, such as models with a significant contribution to reionization from active galactic nuclei (AGN; \citealt{volonteri09}).  The bright end of the AGN luminosity function declines drastically between $z=3$ and $z=6$ \citep{fan01}, precluding rare AGN (i.e., quasars) as the primary driver of reionization.  Nevertheless, reionization by fainter AGN may still be possible (\citealt{siana08, shankar07, glikman11}, although see \citealt{willott10}).  In addition, at least a fraction of reionization could have been sourced by soft X-rays produced from inverse Compton scattering of cosmic microwave background (CMB) photons off supernova-accelerated electrons \citep{oh01}, by emission produced directly in supernova remnant shocks \citep{johnson11}, by high-mass X-ray binaries \citep{power09, mirabel11}, by mini-quasars \citep{madau04}, by halo accretion shocks \citep{dopita11, wyithe11}, or by dark matter annihilations \citep{belikov09}.  Even if stars reionized the bulk of the intergalactic hydrogen, \citet{oh01}, \citet{johnson11}, and \citet{mirabel11} argued that X-rays associated with the remnants of these stars should contribute $\sim 10\%$ of the ionizations via any of a few mechanisms.

X-ray reionization scenarios have recently received renewed attention (e.g., \citealt{haimanopinion}), particularly for two reasons.  Firstly, soft X-ray photons would have had little difficulty in escaping from $z\sim 10$ galaxies and ionizing the intergalactic medium (IGM).  This is in contrast to EUV photons, for which it is unclear whether they would have escaped at all (e.g., \citealt{gnedin08}).  Empirically, observations of $z\sim 3-4$ Ly$\alpha$ emitters suggest that a non-negligible fraction of ionizing photons was escaping into the IGM \citep{nestor11}.  Curiously, for Lyman-break galaxies the escape fraction is constrained to be smaller.  \citet{vanzella10} placed an upper bound on the escape fraction of $3.4 <z<4.5$ Lyman breaks of $<5-20\%$.  Secondly, there is evidence that high-mass X-ray binaries (HMXBs) and ultra-luminous X-ray sources (ULXs) anti-correlate with metallicity \citep{crowther10, kaaret11}.  This anti-correlation suggests that these X-ray sources -- which already dominate the X-ray production of low-redshift star-forming galaxies -- were even more prevalent at high redshifts \citep{mirabel11}. 

A full or partial reionization by soft X-rays would have resulted in a different morphology of intergalactic \HII\ regions during this process compared to a full reionization by EUV photons alone.  The mean free path of soft X-ray photons is much longer than EUV photons owing to the strong energy dependence of the photoionization cross section.  A photon with energy $E_\gamma$ has a mean free path, $\lambda_{\rm HI}$, to be absorbed by hydrogen of
\begin{equation}
\lambda_{\rm HI} \approx 7 \, x_{\rm HI}^{-1} \, \left( \frac{E_\gamma}{200 {\rm ~eV}} \right)^{2.6} \left( \frac{1+z}{10} \right)^{-2}~{\rm cMpc},
\label{eqn:LHI}
\end{equation}
where $x_{\rm HI}$ is the fraction of hydrogen that is neutral.\footnote{The mean free path to be absorbed by \HeII\ and \HeI\ is comparable, with $\lambda_{\rm HeI} = 4 \, x_{\rm HeI}^{-1}~$cMpc and $\lambda_{\rm HeII} = 5 \, x_{\rm HeII}^{-1}~$cMpc, assuming $E_\gamma = 200$~eV and $z=9$.  $\lambda_{\rm HeI}$ has a softer scaling with $E_\gamma$ than $\lambda_{\rm HI}$ such that the \HeI\ would be the first species to be ionized by a background of just soft X-ray photons.}
If a significant fraction of the IGM were reionized by $E_\gamma \gtrsim 200~$eV photons, the inhomogeneous structure on scales of $\sim 10$~comoving~Mpc that is anticipated by most models of this process would have been largely erased.  A more homogeneous reionization scenario than that preferred by models may be more consistent with limits on the kinetic Sunyaev-Zeldovich anisotropy from reionization \citep{reichardt11, mesinger11}.  High-redshift X-ray production is also important for the reheating of the Universe and determines whether redshifted 21cm radiation appears in emission or absorption \citep{furlanettoohbriggs, mcquinn12}.

This study investigates the constraints from current data on high-redshift X-ray emissions.  Because the Universe is optically thin to $\gtrsim 1~$keV photons from $z\sim 10$, such photons produced during the reionization era would have streamed freely and altered the ionization state of intergalactic metals at $z\sim 3$ as observed in quasar absorption spectra.  They also would contribute to the observed soft X-ray background \citep{dijkstra04, salvaterra07}.  In addition, the hardness of the high-redshift EUV and X-ray emissions determines whether the electrons of helium were ionized concurrently with those of the hydrogen or well after.  An early reionization of the second electron of helium appears to be in conflict with the mounting evidence that helium reionization ended at $z\approx 2.7$ \citep{theuns02, hui03, furlanettodixon, mcquinnGP, shull10, becker11, worseck11}.  Finally, measurements of the thermal state of the IGM have been extended to $4 <z< 6.5$ in the past few years \citep{becker11, bolton11}, redshifts more sensitive to the heating during hydrogen reionization.  Reionization by a harder spectrum than stars would have injected more heat into the IGM, potentially resulting in higher temperatures.

Section \ref{sec:sources} summarizes the leading candidates for sourcing the production of high-redshift soft X-ray and EUV photons.  Section \ref{sec:constraints} discusses the constraints on high-redshift X-ray production from the unresolved soft X-ray background, from $z\sim 2.5$ metal absorption line observations, from our knowledge of the reionization history of the hydrogen and helium, and, finally, from constraints on the thermal history of the IGM.  In our calculations, we assume a flat $\Lambda$CDM cosmological model when necessary with $\Omega_m=0.27$, $h=0.71$, $\sigma_8= 0.8$, $n_s=0.96$, and $\Omega_b = 0.046$, consistent with recent measurements \citep{larson11}.

\section{Potential X-ray Sources at High-redshifts}
 \label{sec:sources}

Here we provide a brief synopsis of source populations that studies have argued could dominate the production of X-rays at high-redshifts and ionize $\gtrsim10\%$ of the Universe.
\begin{description}
\item[AGN:]  The abundance of the brightest AGN is observed to fall off rapidly with redshift at $z>3$ \citep{fan01}.  However, the abundance of AGN that lie on the faint end of the luminosity function at $z>3$ is not as well constrained and could conceivably have grown with increasing redshift (although see \S \ref{ss:direct}).  The average spectral index of the intensity per unit frequency for low-redshift Type I quasars is measured to be $\alpha = -1.7$ in the X-ray \citep{tozzi06} and is found to have a similar index in the far UV \citep{telfer02}.\footnote{If just $1\%$ of the energy to grow into a black hole of mass $10^5~M_\odot$ goes into ionizations and there is $1$ black hole for every $10$ cubic cMpc of volume, black holes would have been able to reionize the Universe.  (High-redshift quasar luminosity function measurements are only sensitive to AGN with $>10^7~M_\odot$; \citealt{willott10b}).  This mass density in massive black holes is a factor of $25$ times smaller than the SMBH density at $z=0$ \citep{yu02} and is consistent with the conservative theoretical estimates in \citet{volonteri08}.  In addition, if there were a relation that tied black hole mass to the total stellar mass in a galaxy at high redshifts (similar to the Magorrian relationship at lower redshifts), \citet{lidz07} showed that black holes would have produced a comparable time--integrated number of ionizing photons to stars.}
\item[The end products of stellar evolution:] Previous studies have suggested several possibilities that fit within this class:
\begin{enumerate}
\item {\it from HMXBs --}   Emission from accretion onto the primary -- a black hole or neutron star -- dominates the X-ray output of low-redshift star forming galaxies.  \citet{mirabel11} argued that these sources are likely to be more abundant per unit of star formation at high redshifts and, thus, to have contributed to the reionization of the Universe.  The population--averaged spectral index for HMXBs has been measured to be $\alpha \approx -0.7$ at $1-10~$keV \citep{swartz04, rephaeli95}, but it is unknown whether this power-law continues to lower energies.
\item {\it from the cooling of supernova-accelerated electrons --}  At high redshifts, electronic cosmic rays that were accelerated in supernovae would have cooled primarily by inverse Compton scattering off the CMB \citep{oh01}.   \citet{oh01} argued that this mechanism could naturally yield $10\%$ of the ionizing photons produced by a stellar population if $f_{\rm esc} \approx 0.01$, where $f_{\rm esc}$ is the fraction of $1~$Ry photons that can escape from within galaxies.  This mechanism generates a spectrum with spectral index in $I_E$ of $-\gamma/2$, where $\gamma$ is the input power-law of the electron spectrum.  It is measured to be $\approx 2$ for the injected cosmic ray spectrum in the Milky Way halo \citep{strong04}.  We subsequently assume $\gamma =2$.\footnote{The fiducial values for X-ray production in \citet{oh01} are likely optimistic.  Estimates for the fraction of the ejecta energy that goes into accelerated electrons find $0.1-1\%$ rather than the $10\%$ value used in this study (e.g., \citealt{kobayashi04, thompson06}).}

\item {\it from shocks in supernova remnants --}  \citet{johnson11} estimated that direct emission from such shocks  produces $\approx 10\%$ of the ionizing photons that originate from a stellar population.   Using models of shock emission, we find that this process results in a time-integrated spectrum that can be approximated with a spectral index of $\alpha \approx -1.7$ at $< 1~$keV and with a falloff above this energy.\footnote{This calculation uses the spectral libraries from supernovae shock calculations of \citet{allen08}, averaging over shock velocities of $<1000~$km~s$^{-1}$ as outlined in \citet{johnson11}.} 

\end{enumerate}
\item[Dark matter annihilations:] The electronic by-products cool by inverse Compton (or synchrotron) emission.  This cooling results in a steady-state e+e- spectrum with equal energy distributed per log energy up to energies of order the particle mass and an inverse Compton spectrum in $I_E$ with spectral index $-1/2$ \citep{mcquinnHAZE}.  \citet{belikov09} found that a generic weakly interacting massive particle could result in an ionized fraction of $10^{-2}$ and that recently popular models with an enhanced annihilation cross section could even result in a full reionization. 

\end{description}
Some of the aforementioned sources could be partially obscured by a column of hydrogen and helium gas, which would further harden their ionizing emission.\footnote{A hydrogen column of $N_{\rm HI}$ has a photoionization optical depth of unity to photons with $E \lesssim 1 \, (N_{\rm HI}/{10^{23}~ {\rm cm^{-2}}})^{0.35}$~keV.   Significant obscuration impacts $75\%$ of AGN \citep{tozzi06} and many HMXBs \citep{lutovinov05}.}

\subsection{Direct Observations of UV and X-ray Sources}
\label{ss:direct}
Measurements of the quasar optical luminosity function as well as of X-ray source counts place constraints on the abundance and evolution of the aforementioned sources.   The standard lore is that the quasar luminosity function falls off quickly towards high redshifts such that quasars were a subdominant contribution to the reionization of the IGM (e.g., \citealt{madau99, faucher08b}).  However, this conclusion has stemmed from observations of bright, $L>L_\star$ quasars at $z\geq3$.

Recent studies have begun to constrain the faint-end slope at $z>3$ \citep{dijkstra06, siana08, shankar07, willott10, glikman11}.   Of note, \citet{willott10} placed an order of magnitude more stringent constraint on the abundance of faint $z\sim6$ quasars than other analyses, using the Canada-France High-z Quasar Survey.  With an absolute magnitude limit of $M_{1450{\rm A}} = -22$ plus extrapolations to fainter magnitudes, \citet{willott10} found that quasars could contribute no more than $10\%$ to the ionizations required to maintain reionization at $z=6$.  If this result is confirmed and there is not a significant turn-up in the quasar abundance towards fainter luminosities, quasars alone cannot reionize the Universe.\footnote{We note that all estimates for the quasar ionizing photon emissivity use a single template to extrapolate from the source-frame far UV to the source-frame EUV.  This extrapolation would be in error if spectral hardness anticorrelates with ultraviolet luminosity, as has been suggested in \citealt{scott04}.}  In this paper, we will remain agnostic as to whether AGN reionization scenarios are allowed and investigate whether they can also be constrained with other techniques.

At the faintest fluxes measured with the Chandra Space Telescope, star-forming galaxies become an important contribution to the observed X-ray source counts \citep{hickox07}.  
In fact, if the faint end of the quasar luminosity function does indeed decline with increasing redshift at $z>3$, the dominant high-redshift X-ray sources in the Universe were likely star-forming galaxies.  There are some constraints on how the X-ray emissions from galaxies evolve with redshift. \citet{nandra02} and \citet{reddy04} found that stacked $z= 1.5-3$ Lyman-break galaxies 
are consistent with the hard X-ray luminosity--star formation rate relationship found at $z=0$ \citep{reddy04}.  \citet{cowie11} reached a similar conclusion for galaxies at redshifts of $1<z<6$.  They found a factor of $3$ higher normalization than the $z=0$ relation of \citet{mineo11} at all redshifts (but similar to the $z=0$ determination in \citealt{dijkstra11}), with a factor of few uncertainty on their determination at $z>4$.   It is worth noting that these constraints are on star formation in the most luminous high-redshift galaxies, which may not be representative of most coeval star formation.  

\section{Indirect Constraints on X-ray Source Models}
\label{sec:constraints}

The following investigates additional constraints on these X-ray source models from measurements sensitive to the total amount of emission (in contrast to luminosity function measurements, which often probe only the brightest contributors).

\subsection{The Soft X-ray Background}
\label{sec:SXB}

This section revisits the constraints on X-ray reionization scenarios from observations of the soft X-ray background (SXB).  Such constraints were first discussed in \citet{dijkstra04}, and an aim of this section is to update these calculations.  The soft X-ray background is one of the most constraining backgrounds for $z\sim 10$ sources because it is not altered by intervening absorption, in contrast to optical and ultraviolet backgrounds.  Our focus is on the constraints from reionization scenarios in which only a few ionizing photons produced per hydrogen atom were required to complete reionization, a different region of parameter space than originally studied in \citet{dijkstra04}.   Both observations of the $\sim 1~$Ry emissivity of sources from the Ly$\alpha$ forest at lower redshifts and investigations of the expected number of recombinations during reionization suggest that reionization was in this regime  \citep{miralda03, bolton07, pawlik09, mcquinn11}.  

The SXB can be used to constrain the comoving number density of ionizing photons, $n_i$, emitted before the end of reionization.  If this number density is greater than the number of hydrogen atoms, then there were enough photons to reionize the Universe.  In particular, in the absence of foreground emission, $n_i$ is related to the intensity of the $z=0$ background per unit energy, $I_E$, via 
\begin{equation}
n_i = \frac{4\,\pi}{c} \int_{13.6 \,{\rm eV}/(1+z_{\rm E})}^{E_{\rm max}/(1+z_{\rm E})} dE \, \frac{I_E}{E} \exp[-\tau_{\rm eff}(E, z_E)],
\label{eqn:ngamma}
\end{equation}
where $\exp[-\tau_{\rm eff}(E, z_E)]$ is the average transmission for a photon that travels from redshift $z_E$, reaching the Earth with energy $E$, and here we have assumed for simplicity a single redshift $z_E$ for the emission.  Our subsequent calculations will assume $E_{\rm max} = 1.8 \, [(1 + z)/15)]^{0.5} \,x_{\rm HI}^{1/3}~$keV, where $x_{\rm HI}$ is the \HI\ fraction, which we take to be $0.5$.  This corresponds to the photon energy that has a photoionization optical depth equal to unity to travel a Hubble distance at this neutral fraction and redshift.  However, our results depend weakly on $E_{\rm max}$.

X-ray photons originating at $z\sim 10$ with energy $> 1~$keV have a negligible probability to be absorbed (i.e., $\exp[-\tau_{\rm eff}] = 1$).  Thus, since $I_E$ is measured in the X-ray, if we make assumptions for how $I_E$ scales to lower $E$, we can constrain $n_i$.  We adopt $3.5\,\times10^{-13}~$erg~cm$^{-2}$~s$^{-1}$~deg$^{-2}$ for the unresolved soft X-ray background intensity between $0.5$ and $2~$keV.  This value was derived in \citet{moretti03} and \citet{dijkstra04} by combining both widefield and pencil-beam X-ray surveys and also by subtracting off the small contribution owing to Thomson scattered radiation from known sources.  This background estimate is consistent with the estimate of \citet{hickox07} from the Chandra deep fields, which was $3.5\,(\pm1.4)\times 10^{-13}~$erg~cm$^{-2}$~s$^{-1}$~deg$^{-2}$ between $1$ and $2~$keV.  (We are assuming that all of the resolved X-ray sources are from after reionization.  This is likely the case.  In fact, the \citealt{hickox07} measurement masked point sources detected with the Hubble Space Telescope in the Z-band, which incidentally would have missed masking sources at $z\gtrsim 6.5$ owing to \HI\ Lyman-series absorption.)

For a spectral index of $\alpha$ for the sources' intensity per unit frequency, the number of ionizing photons produced per baryon $N_i (\alpha) \equiv n_i/n_H$ if \emph{all} of the unresolved background owes to reionization era sources is
\begin{eqnarray}
N_i(\{-0.5, -1, -1.5\})  &\approx&   \{0.05, 0.8, 14\} ~I_{3.5}\label{eqn:Ni}\\
 &&\times \left(\frac{1 +z_E}{10}  \right)^{\{-0.5, -1, -1.5\}},  \nonumber
\end{eqnarray}
where $I_{3.5}$ is the $0.5-2~$keV intensity of the soft X-ray background in units of $3.5 \times10^{-13}~$erg~cm$^{-2}$~s$^{-1}$~deg$^{-2}$, and $z_E$ is the emission redshift.
Even though we have assumed that all the photons were emitted at a single redshift, the results are rather weakly sensitive to this choice for reionization era sources.  Equation~(\ref{eqn:Ni}) demonstrates that the SXB can rule out X-ray reionization models that satisfy $\alpha \gtrsim -1$ as these have $N_i <1$.

\begin{figure}
\begin{center}
{\epsfig{file=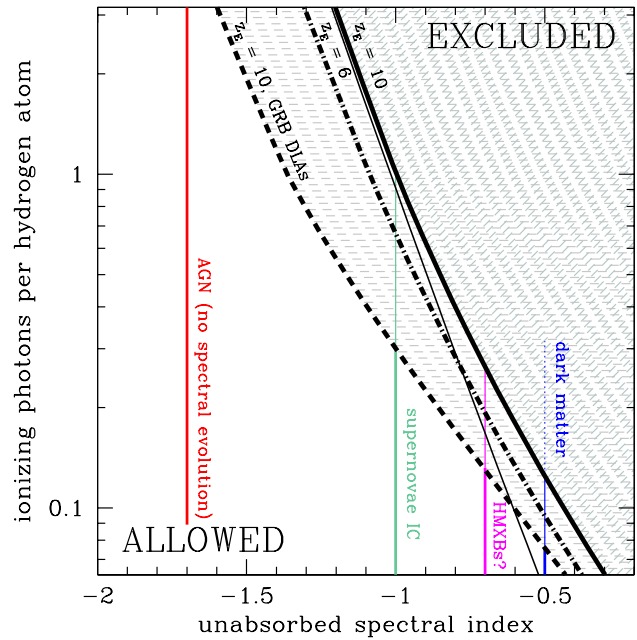, width=9cm}}
\end{center}
\caption{Constraints from observations of the unresolved soft X-ray background (SXB) intensity on the ionizing photon production of high-redshift sources.  Shown is the number of such photons per hydrogen atom a source population could have produced as a function of its effective spectral index in $I_E$ between $1~$Ry and the hard X-ray band, denoted as $\alpha$ in the text.  The solid curves (dot-dashed curve) are the contours that saturates the unresolved SXB for the case where the photons were produced instantaneously at $z=10~$(6).  The thin solid curve omits secondary ionizations, while the the other curves account for them as described in the text.  For the scenario represented by each curve, the parameter space above its curve is excluded.   The ``GRB DLAs'' curve is the same as the thick solid curve but the photons are assumed to have been filtered by the host-galaxy \HI\ column density distribution inferred from GRB afterglows (which models interstellar absorption for galactic sources).  The labeled vertical lines show the fraction of ionizations that could result from the different source populations discussed in \S \ref{sec:sources}, using empirically motivated estimates for their power-law index.   
\label{fig:UVB_const}}
\end{figure}

Figure \ref{fig:UVB_const} illustrates in more detail the upper bound from the unresolved SXB on $N_i$ as a function of $\alpha$.  The solid curves (dot-dashed curve) are for the case in which all of the photons were produced at $z=10~$(6), assuming no absorptions from within the progenitor galaxy.  The thin solid curve does not include secondary ionizations, while the three thick curves include them (which is the more realistic case).  Secondary ionizations, which were ignored in the previous discussion, are included by assuming that the ionizations occurred in a medium that was initially neutral, ignoring recombinations, and using the secondary ionization rates in \citet{ricotti02}.  (Even during a patchy reionization scenario, X-ray ionizations typically would have occurred outside of the \HII\ regions that surround EUV sources where the IGM was 
largely neutral.)  This approximation allows us to map the previous estimates for $n_i$ (eqn. \ref{eqn:Ni})  to a boosted $n_i$ that includes secondaries.  The differences between the thick and thin solid curves demonstrate that secondary ionizations only impact $N_i$ for $x_i \ll 1$.  Otherwise, the excess energy per ionization is channeled primarily into heating the IGM rather than into secondary ionizations \citep{shull85}.  

The dashed curve labeled ``GRB DLAs'' in Figure \ref{fig:UVB_const} is the same as the solid curve but where the emitted ionizing photons are filtered by the host-galaxy \HI\ column density distribution inferred from gamma ray burst (GRB) afterglow spectra.  It is plausible that this distribution is the same as the column-density distribution to escape galaxies from star forming regions \citep{gnedin08b}.  This GRB column-density distribution results in $f_{\rm esc} \approx 0.02$ for $1~$Ry photons \citep{chen07}.  The labeled vertical lines in Figure \ref{fig:UVB_const} represent the spectral index expected for the sources discussed in \S \ref{sec:sources}.  The constraints on AGN are weak, which agrees with the conclusion of \citet{salvaterra07} that AGN reionization scenarios can be consistent with the unresolved SXB.   We have conservatively assumed that the average spectral index of AGN is $-1.7$ as found for quasars at lower redshifts; either obscuration or a hardening of the spectrum with redshift (potentially stemming from the emission originating from lower mass central black holes) 
 would tighten these bounds.  The constraints from inverse Compton from supernova-accelerated electrons are more significant.  This mechanism could not fully reionize the \HI\ and is limited to ionizing $\lesssim 30\%$ of the Universe in the case with GRB damped Lyman-$\alpha$ systems (DLAs).  The same conclusion holds for any process that results in a steady state number of electrons per unity energy with a power-law index of $-3$ in energy or harder.  The steady-state electron distribution would be $-2$ for the electronic byproducts of dark matter annihilations in standard weakly interacting massive dark matter particle models (Section \ref{sec:sources}).  In this case, the constraints on $N_i$ are even tighter ($N_i \lesssim 10\%$);  however, in this case the approximation of the sources turning on and off at a single redshift cannot hold, and the associated ionizing emission is likely to also occur at lower redshifts.  Metal absorption line measurements rule out a significant contribution from X-ray sources that are still active at $z=3$ with $\alpha > -1$ (\S \ref{sec:EUVB}).  If we use the estimate for the spectral index for the galaxy-averaged emission of HMXBs and from ULXs of $\alpha = -0.7$ \citep{rephaeli95, swartz04}, then HMXBs/ULXs can also only contribute $\lesssim 10-20\%$ of the ionizations.  If their spectrum were harder than this index, as is likely at lower frequencies (especially since this emission is largely thermal radiation from a truncated accretion disk with characteristic temperatures of $T\sim 1-10~$keV), then the limit on their contribution would have been even tighter.  
 
 We note that all of the aforementioned constraints improve proportionally to the measured unresolved SXB intensity.  It is quite likely that a large fraction of the unresolved X-ray background was produced after reionization by galaxies below the current flux threshold of the observations, by the local bubble, or the warm hot intergalactic medium.  In future missions, lower frequencies (that are closer to the EUV band) would be interesting to target with SXB measurements, as we estimate that the foreground extragalactic \HI\ absorption for photon energies $>200~$eV has effective optical depth of less than unity (and the optical depth contributed by the Milky Way is also less than unity at $>200~$eV).

\begin{figure}
\begin{center}
{\epsfig{file=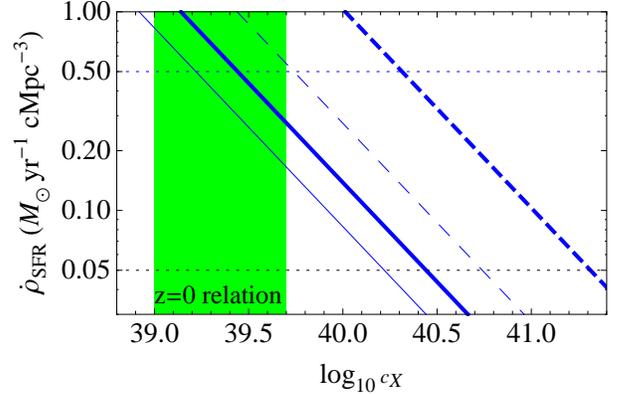, width=8cm}}
\end{center}
\caption{The $2-8~$keV luminosity produced per unit star formation rate, $c_X$, required to saturate the unresolved SXB, for different cosmic star formation rate densities, $\dot \rho_{\rm SFR}$.  $c_X$ is given in units of erg~s$^{-1}$/($M_\odot$~yr$^{-1}$~cMpc$^{-3}$).  The curves show the values which saturate the unresolved SXB under the assumption that the sources radiated with a constant emissivity
from $z = \infty$ to either $z = 3$ (solid curves) or to $z=8$ (dashed curves). The thick curves are for $\alpha = -1$, and the thin are for $\alpha= 0$.  The dotted horizontal lines denote the star formation rate density required to maintain reionization at $z=6$ for $f_{\rm esc} = 0.1$ (lower horizontal line) and $f_{\rm esc} = 0.01$ (upper horizontal line; see text for details).   The shaded region is the constraint on the relationship between X-ray luminosity and star formation rate from $z=0$ observations \citep{mineo11}.  Its width is set by the observed scatter about the mean relation.
\label{fig:SFR_const}}
\end{figure}

Figure \ref{fig:SFR_const} considers the type of evolution that is necessary for galactic X-ray emission (probably associated with HMXBs) to saturate the unresolved SXB intensity.\footnote{The calculations shown in this figure are similar to those in \citet{dijkstra11} except that we do not tie our constraints to the observed $\dot \rho_{\rm SFR}(z)$, which suffers from incompleteness at high redshifts.}  The curves represent the $2-8~$keV luminosity per unit star formation rate, $c_X$, needed to saturate the SXB for different cosmic star formation rate densities, $\dot \rho_{\rm SFR}$.  The curves assume that the sources radiate at a constant emissivity from $z=\infty$ -- but the dependence on the maximum redshift is weak -- to $z= 3$ (solid curve) and $z=8$ (dashed curve).  The thick curves are for $\alpha = -1$, and the thin for $\alpha = 0$.  The lower and upper horizontal lines denote the star formation rate (SFR) for a Salpeter initial mass function required to maintain reionization (balance recombinations) for a clumping factor of $5$ at $z=7$ for $f_{\rm esc} = 0.1$ (requiring  $\dot \rho_{\rm SFR} \approx 5\times 10^{-2}~$$M_\odot$~yr$^{-1}$~cMpc$^{-3}$) and $f_{\rm esc} = 0.01$ (requiring  $\dot \rho_{\rm SFR} \approx 5\times 10^{-1}~$$M_\odot$~yr$^{-1}$~cMpc$^{-3}$) \citep{madau99}.     Estimates from the Hubble Ultra Deep Field fall on the low side of these $\dot \rho_{\rm SFR}$, favoring $\dot \rho_{\rm SFR} \approx 10^{-2}~$$M_\odot$~yr$^{-1}$~cMpc$^{-3}$ at $z=8$ and an order of magnitude larger value at $z=3$ \citep{bouwens11b}, but such estimates miss fainter galaxies which likely dominate $\dot \rho_{\rm SFR}$, especially at $z\gtrsim 6$ \citep{kuhlen12}.  The shaded region is the constraint from \citet{mineo11} on $c_X$ at $z = 0$, with its width set by the the observed scatter in this relation.

For the highest values of $\dot \rho_{\rm SFR}$ that are considered Figure \ref{fig:SFR_const} (which would require significant incompleteness in current SFR measurements), not much evolution with redshift would be required to saturate the bound from the SXB.  For the lower values of $ \dot \rho_{\rm SFR}$, however, an order-of-magnitude increase in $c_X$ or more would be needed.  At least for the galaxies with the highest SFRs, such significant evolution is marginally ruled out (\citealt{cowie11}; Section \ref{ss:direct}).

\subsection{Quasar Absorption Lines}
\label{sec:EUVB}

\HI\ Ly$\alpha$ forest and metal line absorption measurements from quasar spectra have been measured to redshifts as high as $z\approx 7$ (\citealt{mortlock11}; see \citealt{meiksin09} for a recent review of these diagnostics).  These measurements are sensitive to the coeval soft X-ray background.  Unlike the $z=0$ unresolved SXB, constraints derived from high-redshift absorption lines avoid contamination from lower redshift, faint emission.  However, existing measurements of quasar absorption lines at $z>3$ do not unambiguously constrain these radiation backgrounds as typically few lines are detected for each absorption-line complex (in part because the Ly$\alpha$ forest is so saturated).  Therefore, we concentrate on the constraints from $z\approx 2.5$ -- the ``sweet spot'' for quasar absorption studies.

\begin{figure}
\begin{center}
{\epsfig{file=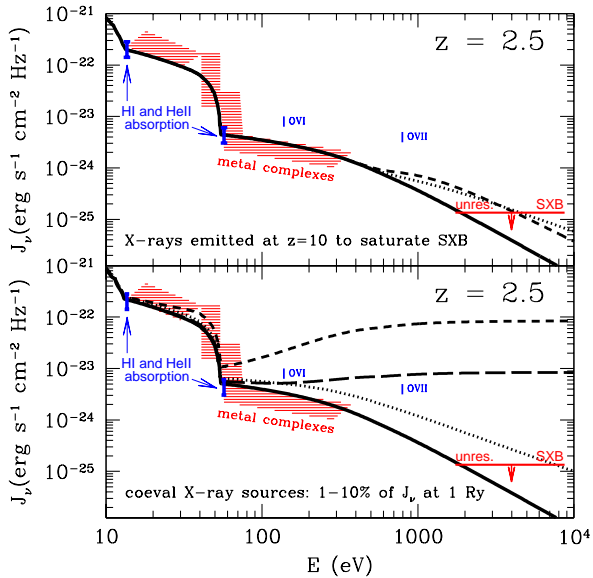, width=8cm}}
\end{center}
\caption{Effect of hard sources on the EUV and X-ray backgrounds at $z=2.5$, as well as constraints from the \HI\ and \HeII\ Lyman forests and from the metal complexes studied in \citet{agafonova07}.  The fiducial background model discussed in the text in which quasars dominate the emissivity is given by the black solid curve in both panels.  The other models in the top panel assume that an extra component of the emission was released at $z=10$ and such that it saturates the $z=0$ unresolved SXB intensity with $\alpha = 0$ (dashed) and $\alpha = -1$ (dotted).  Also shown is the upper bound for the $z=0$ unresolved SXB under the most conservative assumption that all contributing emissions originated from $z>2.5$.  The other models in the bottom panel assume that $10\%$ or $1\%$ of hydrogen ionizing photons originate from a harder source than quasars.  As in the top panel, the dashed curves assume $\alpha = 0$ and the dotted curves assume $\alpha = -1$ (shown only for the $10\%$ case).  Lastly, the ionization potentials of \OVI\ and \OVII\ are marked. \OVI\ is the metal line that is sensitive to the highest energies for analyses at $z=2.5$.
\label{fig:spectrum_const}}
\end{figure}

\HI\ and \HeII\ Ly$\alpha$ forest measurements at $z\approx 2.5$ constrain the the \HI\ and \HeII\ photoionization rates, $\Gamma_{\rm HI}$ and $\Gamma_{\rm HeII}$.  We define $\eta$ as the local ratio of the \HI\ to \HeII\ column density, which is proportional to $\Gamma_{\rm HI}/\Gamma_{\rm HeII}$ in photoionization equilibrium.   These quantities can be related to the angular-averaged intensity per unit frequency at $\approx 1$ and $\approx 4~$Ry as
\begin{eqnarray}
J_\nu({\rm 1\, Ry}) &\approx& \frac{(3+\alpha_{\rm bk}) h \, \Gamma_{\rm HI}}{4\pi \sigma_{\rm HI}}, \label{eqn:1Ry}\\
&\approx& 2.5 \times 10^{-22} \left( \frac{3+\alpha_{\rm bk}}{3} \right) \, \left(\frac{\Gamma_{\rm HI}}{10^{-12} {\rm ~s}^{-1}}\right);  \nonumber\\
J_\nu({\rm 4\, Ry}) &\approx& \frac{(3+\alpha_{\rm bk}) h \, \Gamma_{\rm HeII}}{4 \pi \sigma_{\rm HeII}}, \label{eqn:4Ry}\\
&\approx& 6.9 \times 10^{-24} \left( \frac{3+\alpha_{\rm bk}}{3} \right) \frac{80}{\eta} \, \left(\frac{\Gamma_{\rm HI} }{10^{-12}{\rm ~s}^{-1}}\right), \nonumber
\end{eqnarray}
where the second and fourth lines use c.g.s units, $\alpha_{\rm bk}$ is the spectral index of the EUV background, and $\sigma_X$ is the photoionization cross section at the ionization potential of ionic species $X$.  Current measurements find $\Gamma_{\rm HI} \approx 0.5-1\times 10^{-12}~{\rm s}^{-1}$ and $\eta \approx 80$ at $z=2.5$ \citep{bolton05, faucher08, worseck11}.  The error bars at $1$ and $4~$Ry in Figure \ref{fig:spectrum_const} show the constraints on $J_\nu$ using equations (\ref{eqn:1Ry}) and (\ref{eqn:4Ry}) with $\Gamma_{\rm HI} = 0.75 \, (\pm 0.25) \times 10^{-12}~{\rm s}^{-1}$.  

Metal absorption line measurements in quasar spectra also constrain the $z\approx 2.5$ EUV background (e.g., \citealt{agafonova07, fechner11}).  The red shaded trapezoids in Figure \ref{fig:spectrum_const} are the best-fit EUV background spectrum for two $z\approx 2.5$ metal absorption complexes presented in \citet{agafonova07}, where $\approx 12$ absorption lines were simultaneously fit for each complex.  We have specified the normalization of the shaded region, which is unconstrained owing to degeneracy with the level of enrichment.   In addition, the finite width of the shaded regions is for viewing ease and reflects a factor of $\approx 2$ shift in this constraint.   It does not reflect the uncertainty in this measurement as error bars were not provided in \citet{agafonova07}.

To investigate the impact of additional hard sources of emission on the EUV background and how much of such radiation the quoted absorption line constraints allow, we construct models for the $10-10^4~$eV radiation background.
In these models, the mean specific intensity of the meta-galactic radiation background at frequency $\nu_0$ and redshift $z_0$ is (e.g., \citealt{peebles93, haardt96}) 
\begin{equation}
J(\nu_0, z_0) = \frac{\left({1+z_0} \right)^3}{4\pi} \int_{\eta(z_0)}^\infty d\eta  \, a \, \epsilon(\nu, z) \, e^{-\tau_{\rm eff}(\nu_0, z_0, z)}.
\label{eqn:Jnutrans}
\end{equation}
Here, $d\eta = c \,dz/H(z)$ is the differential of the conformal distance, $\epsilon(\nu, z)$ is the specific comoving emissivity at a frequency of $\nu \equiv \nu_0 (1+z)/(1+z_0)$, $\exp[-\tau_{\rm eff}]$ is the probability that a photon is transmitted as it travels between $z$ and $z_0$, and
\begin{equation}
\tau_{\rm eff} = \int_{z_0}^z dz \int_0^\infty dN_{\rm HI} \, \frac{d{\cal N}}{dzdN_{\rm HI}} \, \left(1- e^{-\tau(N_{\rm HI})} \right).
\label{eqn:taunutrans}
\end{equation}
$d{\cal N}/dzdN_{\rm HI}$ is the \HI\ column density distribution (constrained by quasar absorption-line measurements), and $\tau(N_{\rm HI})$ is the \HI\ and \HeII\ continuum optical depths of a system with column $N_{\rm HI}$ at frequency $\nu_0 \,(1+z)/(1+z_0)$.   Equations (\ref{eqn:Jnutrans}) and (\ref{eqn:taunutrans}) are exact in the limit of uncorrelated (Poissonian) absorbers.

We parameterize $d{\cal N}/{dzdN_{\rm HI}}$ as a power-law with index $-1.5$ (which is simplistic; \citealt{prochaska09}, but our qualitative results are weakly sensitive to this choice).\footnote{We assume $d{\cal N}/{dzdN_{\rm HI}}$ is normalized so that the the mean free path of $1~$Ry photons is $300~$cMpc at $z=3$ and so that it decreases as $(1+z)^{-2}$ with increasing redshift, in agreement with measurements \citep{prochaska09, prochaska10}.  Furthermore, we assume that the average diffuse \HeII\ fraction is given by $\tanh([z-3]/1.5)$, although the results depend weakly on the assumed \HeII\ ionization history because the frequency that has optical depth equal to $1$ scales as the \HeII\ fraction to the one-third power.  After \HeII\ reionization and in \HeII\ bubbles, we assume a $4~$Ry photon has mean free path of $50~$cMpc at $z=3$, which is consistent with other estimates (e.g., \citealt{faucher09, haardt11}).\\ \\
Our simple background model does not include the emission from Type~2 quasars, which contribute 50-75\% to the resolved X-ray background at $\gg1~$keV (see \citealt{haardt11}).  Accounting for unresolved Type~2 quasars would make it more difficult for these absorption line measurements to accommodate additional hard X-ray emission.}  In addition, we assume quasars with a spectral index of $-1.7$ dominate the emissivity, contributing a comoving emissivity at $1~$Ry of $5\times10^{24}~$erg~s$^{-1}$~Hz$^{-1}$~cMpc$^{-3}$ at $z=3$, and where this emissivity scales as $(1+z)^{-2}$ with redshift, in approximate agreement with estimates for the contribution of AGN at $z> 2.5$ \citep{hopkins07}.  (We ignore starlight, which may contribute as well; e.g. \citealt{faucher08}).

The top panel in Figure \ref{fig:spectrum_const} shows the impact of hard sources on the EUV and X-ray backgrounds at $z=2.5$.  The black solid curve in both panels is the fiducial background model that assumes only quasars.  The other two background model curves in the top panel assume that additional emission was released at $z=10$ with $\alpha = -1$ (dashed curve) and $\alpha = -1.5$ (dotted curve), with its amplitude set such that it saturates the unresolved SXB.  The red upper bound is the \citet{moretti03} constraint on the SXB (i.e., the $z=0$ unresolved $0.5-2~$keV SXB multiplied by $[1+z]^3$ and blueshifted appropriately).  The minimum energy that is impacted by intergalactic absorption, $\sim 1~$keV, is weakly model dependent, as it scales with the amount of hydrogen and helium in our models to the one-third power.  Figure \ref{fig:spectrum_const} shows that high-redshift emission at energies less than $\lesssim 1~$keV -- energies at which absorption line measurements are most sensitive -- is absorbed out by intervening gas by $z = 2.5$.   Thus, the top panel in Figure \ref{fig:spectrum_const} demonstrates that the $z=0$ unresolved SXB provides tighter constraints on relic radiation backgrounds from $z\gg 2.5$ than $z=2.5$ quasar absorption line measurements since it probes energies that are unabsorbed.

The bottom panel in Figure \ref{fig:spectrum_const} shows the $J_\nu$ in the fiducial background model (solid curve) and for cases in which an additional $10\%$/$1\%$ of hydrogen ionizing photons at $1~$Ry originate from a hard source with $\alpha = 0/-1$ (dashed curves/dotted curve, shown only for $\alpha=0$).  These cases are all clearly ruled out by the SXB measurement:  The SXB plus absorption line measurements strongly limit any coeval hard emission.  We have assumed in the bottom panel that the total emissivity of the hard sources falls off in the same manner as is assumed for the quasars, $(1+z)^{-2}$.  However, we find that the qualitative results are unchanged if their emissivity were instead constant with redshift such that the $10\%$/$1\%$ contributions to $J_\nu$ at $1~$Ry would only have been satisfied at $z=2.5$.

In conclusion, the SXB places stronger constraints on hard high-redshift emission than $z\sim 3$ metal absorption lines under the assumption of power-law sources.  Foreground absorption filters out high-redshift emissions at frequencies at which absorption measurements are the most sensitive.  In conjunction, absorption diagnostics and the unresolved SXB show that a source population which contributes significantly to the ionization state of the intergalactic hydrogen and helium at $z\sim 2.5$ must also have a spectrum which cannot be much harder than that of low-redshift quasars.

\subsection{The Reionization History}
\label{sec:HeII}

\begin{figure}
\begin{center}
{\epsfig{file=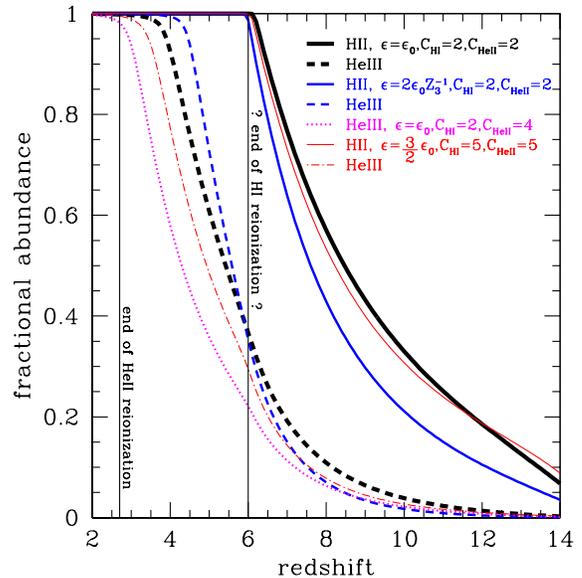, width=8cm}}
\end{center}
\caption{Exploration of whether quasars can ionize both the hydrogen and helium and still be consistent with constraints on the cosmological reionization history.  Shown are the average cosmological ionization fraction in \HII\ and \HeIII\ for reionization with a quasar-like spectrum ($\alpha = -1.7$), under different assumptions for the population's emissivity evolution and for the clumpiness of intergalactic gas.    Here, $\epsilon_0 = 8\times10^{-25}~$erg~s$^{-1}$~Hz$^{-1}$~cMpc$^{-3}$ (as motivated by Ly$\alpha$ forest inferences) and $Z_3 \equiv (1+z)/4$.  The curves whose labels begin as ``\HeIII'' are calculated with the same parameters as the ``\HII'' curves labelled directly above them but instead show the reionization history of \HeIII.  
\label{fig:reion_hist}}
\end{figure}

Constraints on hard high-redshift emission are also set by our knowledge of the cosmological reionization history of hydrogen and helium.  
The Lyman-$\alpha$ forest indicates that the hydrogen in the IGM was significantly ionized by $z\approx6$, when the Universe was $1~$Gyr old (\citealt{fan06}; although, somewhat lower redshifts are still possible; \citealt{mesinger10}).  In addition, there is now significant evidence that \HeII\ reionization was ending at $z = 2.7$ \citep{shull10, mcquinnGP, furlanettodixon, worseck11} -- well after hydrogen reionization and at the time the Universe was $2.5~$Gyr old.  We take the latter as a given for this discussion, noting that Gunn-Peterson troughs are seen in the \HeII\ Ly$\alpha$ forest to redshifts as low as $z = 2.7$ \citep{shull10}.  It is very likely that these indicate large-scale \HeII\ regions \citep{mcquinn09}.  

The Case A \HeIII~$\rightarrow$~\HeII\ recombination time is $\approx 5.5$ times shorter than for hydrogen at relevant temperatures and, at the cosmic mean density, is equal to $0.7$ times the age of the Universe at $z \approx 3$.  Thus, if the sources that doubly ionized the helium shut off at a high redshift, the \HeIII\ would have recombined into \HeII\ by $z=3$ (except in deep voids; \citealt{venkatesan01, mcquinnHeI}).  This shorter recombination time prohibits placing constraints from the observed end redshift of \HeII\ reionization on X-ray source models in which the sources shut off at sufficiently early times.  It does, however, constrain models in which the sources were still active to lower redshifts.  However, we argued in \S \ref{sec:EUVB} that absorption line diagnostics plus the unresolved SXB exclude sources that are still active at $z\sim 3$ from having a harder spectrum than quasars.  Thus, quasars are the most plausible sources this history can be used to constrain.

We aim to calculate ionization histories for different source models.  We define the variable $C_a$ to be the enhancement in the recombination rate into ion $a$ over a homogeneous Universe at a temperature of $10^4~$K.  At relevant redshifts, $C_{a}$ should increase with time (decreasing redshift) as structure forms and as the intensity of the ionizing background within ionized regions increases.

Quasar absorption line measurements themselves place the empirical constraint $C_{\rm HI}<6$ at $z=4$ \citep{miralda03, mcquinn11}.  In addition, recent numerical studies have found $C_{\rm HI}<5$ in cosmological simulations at $z<6$ \citep{mcquinn11} and smaller values at higher redshifts, at least well after reionization \citep{pawlik09}.  These studies also show that this quantity depends weakly on the ionizing background and also on redshift \citep{pawlik09, mcquinn11}.  To estimate $C_{\rm HeII}$ numerically, we have performed a similar calculation to \citet{mcquinn11} in which EUV radiation is shone on all overdense regions within a cosmological simulation volume.  These values yield $C_{\rm HeII} = 1.5-3$ over a large range of photoionizing backgrounds.  We note that since dense regions tend to self-shield more easily at the ionization potential of \HeII\ than at this for \HI, $C_{\rm HeII}$ should always be less than $C_{\rm HI}$.

Next, we take simple AGN emissivity histories that are either constant or power-laws in $1+z$.  The emissivity of $\approx 1~$Ry photons has been measured at $z \approx 2-4$ and found to be just a few ionizing photons per hydrogen atom per Gyr, not significantly in excess of what would have been required to reionize the hydrogen by $z=6$ \citep{miralda03, bolton07, faucher08b}.  Thus, the comoving ionizing emissivity cannot increase significantly from $z\sim 10$ to $z\sim 3$ despite the massive growth of cosmological structures over this interval.

Figure \ref{fig:reion_hist} shows the ionization histories for sources with a quasar-like spectral index of $\alpha = -1.7$.  These curves are calculated by solving for the ionization state of the hydrogen and helium given the attenuated incident spectrum (tracked using many spectral bins).\footnote{We set the electron density to be $(1-Y_{\rm He}) \rho_c(z) \Omega_b/m_p$ plus the electrons from ionic helium, which mimics an inhomogeneous reionization of hydrogen, where $Y_{\rm He} \approx 0.24$ is the \HeII\ fraction and $\rho_c(z)$ is the cosmic critical density.}  The solid lines are the \HII\ fraction as a function of redshift, and the other curves are the \HeIII\ fraction.  The different sets of curves take different clumping factors and emissivity histories, assuming that the sources turning on at $z=15$ and where the emissivity is labeled in terms of $\epsilon_0 \equiv 8\times10^{-25}~$erg~s$^{-1}$~Hz$^{-1}$~cMpc$^{-3}$ (roughly the value that is measured at $z=3$; \citealt{haardt11}).  The red thick curves assume a constant emissivity of $\epsilon_0$ with redshift and $C_{\rm HI}= C_{\rm HeII} = 2$, consistent with recent numerical studies.  In this case, the middle of reionization occurs at $z=9$ (a bit on the low side of what is preferred by WMAP of $10.5\pm1.2$; \citealt{larson11}), \HI\ reionization completes at $z=6$, and \HeII\ reionization ends at $z\approx3.5$.  We require doubling the \HeII\ clumping factor -- which is inconsistent with our estimates -- for \HeII\ reionization to instead end at $z=2.7$ (dotted curve).  In addition, if the emissivity declines with redshift as $(1+z)^{-1}$ from $2 \, \epsilon_0$ at $z=3$, this yields the blue curves, for which \HI\ reionization ends at $z=6$ and \HeII\ reionization ends at $z=4$.  These curves illustrate that the more likely case of a declining quasar emissivity is more difficult to reconcile with \HeII\ reionization ending at $z\approx 2.7$.

In conclusion, it is difficult for faint AGN (or, more generally, sources with $\alpha \geq -1.7$) to be fully responsible for reionization of both hydrogen and helium and to be consistent with the \HeII\ reionization history.  Both hydrogen and helium reionization by faint AGN would require most of the following to hold: (1) that the AGN emissivity does not decline significantly with redshift to $z\sim 10$ in order to maximize the number of recombinations and, thus, the delay before \HeII\ reionization completes, (2) that the typical AGN spectrum does not harden with increasing redshift, (3) that the enhancement in the recombination rate is a factor of a few larger than in recent numerical studies, and (4) that hydrogen reionization ended at the lowest redshifts allowed by observations, $z\approx6$.

\section{The Temperature of the Intergalactic Medium}
 \label{sec:temp}

This final section estimates the values for the intergalactic medium temperature that would have been achieved in plausible reionization models.  Our goal is to understand whether an appreciable fraction of ionizations by soft X-rays would have imprinted a distinct thermal signature.

Photoionization heating is thought to be the dominant heating mechanism for the IGM at high redshifts (shock heating becomes important at $z<2$; \citealt{cen99}), and this heating depends on the sources' properties, both their spectra and luminosities.  Once a region was reionized, it would have been heated to a temperature that depends on the incident ionizing spectrum.  Its subsequent thermal history is determined by processes that are well understood:  Compton cooling off the cosmic microwave background, adiabatic heating and cooling, atomic cooling, and the photoionization heating of the residual neutral gas.  

The timescale for $\lesssim 3\times10^4~$K ionized gas at the cosmic mean density to cool by a factor of $2$ at $z\lesssim 10$ is of the order of the Hubble time. 
Thus, many studies have noted the possibility of doing archeology via $z= 2-6$ temperature measurements \citep{miralda94, hui97, hui03, tittley07, furlanetto09}.  Roughly, temperatures of $>10,000~$K at $z=6$ indicate that part or all of the cosmic gas was ionized within a redshift interval of $\Delta z \approx 3$ from this redshift because Compton cooling efficiently cools the gas that was ionized at $z>9$ below such temperatures \citep{hui03}.  Temperatures of $T >10,000~$K at $z\approx 3$ likely indicate that helium reionization occurred near this redshift as most of the thermal energy from hydrogen reionization has been lost \citep{theuns02, hui03}.  These conclusions depend little on whether hydrogen reionization heated the gas to $15,000~$K or $50,000~$K.  More precise conclusions can be garnered with knowledge of the amount a region was heated to at reionization.  

There have been a few studies that investigated the impact of reionization on the IGM temperature.   \citet{miralda94} argued that if reionization occurred over a significant fraction of a Hubble time, collisional cooling would always result in significant cooling behind the hydrogen ionization front, limiting the temperature that the gas can attain.  We attempt to make this argument more quantitative here.  \citet{tittley07} performed calculations that took into account the inhomogeneous nature of the IGM on the intergalactic temperature during reionization.   However, the planar source model in \citet{tittley07} enhances the prominence of radiative transfer effects owing to inhomogeneities in the IGM and does not treat different front speeds (what we focus on here).  \citet{venkaesan11} investigated the temperature profile in the \HII\ region around a single galaxy, including stellar and black hole emissions, but investigated the heating just immediately around the galaxy and the did not investigate how much a typical parcel of gas was heated during reionization.   Semi-analytic studies of reionization have circumvented knowledge of how much heating occurs at the ionization front by leaving the post-front temperature as a fixed parameter \citep[e.g., ][there $30,000$~K]{furlanetto09}, and these studies followed the subsequent evolution once a region was reionized.    A few radiative transfer simulations of reionization on cosmological scales have also attempted to follow the heating from this process \citep{trac08, finlator11}, but each study investigated a single spectral model and, since they did not resolve the ionization front, did not capture the amount of cooling at the front.  No study has provided a systematic understanding of how the thermal history depends on the spectrum of the sources and the morphology of reionization.  

We have developed a 1D radiative transfer code that performs radiative transfer on the light-cone to calculate the ionization and temperature of the gas around a point source.  This code uses the photoionization cross sections, recombination and collisional ionization coefficients, and cooling rates given in \citet{hui97}.\footnote{We do not employ crude power-law approximations for the cross sections, as is common in similar radiative transfer codes to speed up the computation.}  We also include secondary heating and ionization processes, using the fitting tables from \citet{furlanetto10}, and recombination radiation.  In particular, radiation from recombinations are included in the approximate manner by adding these photons to the outward propagating ray.  We have tested our code against the example calculations in the appendix of \citet{bolton07b}.\footnote{These calculations must resolve the width of the \HI\ ionization front or $(\sigma_{\rm HI} \, n_H) \approx 0.7 \, Z_7^{-3}~$proper kpc, an impossible requirement for a non-adaptive 3D radiative transfer simulation on cosmological scales ($\sim 10~$Mpc).  (See \citealt{cantalupo11} for an adaptive 3D algorithm that can capture this physics.)  In all of the calculations presented here, the front width is resolved with $\approx 10$ elements, which we find is sufficient for convergence.}

 \begin{figure}
\begin{center}
{\epsfig{file=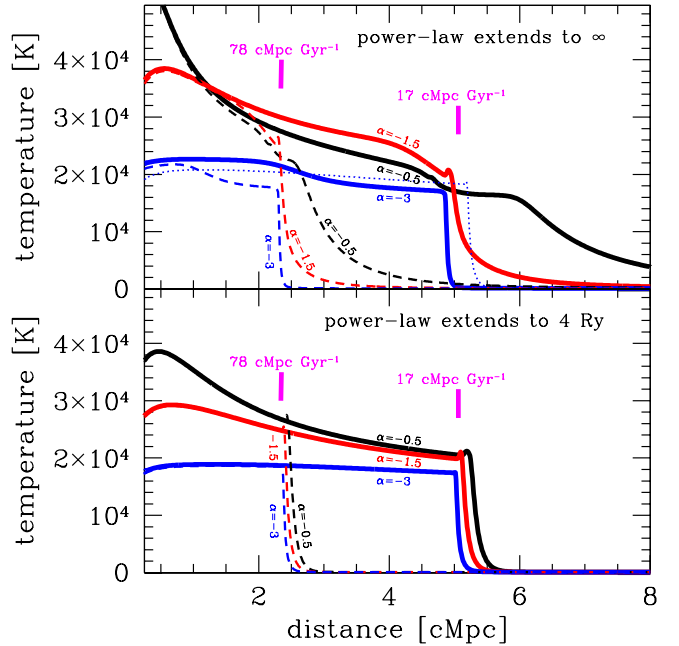, width=9cm}}
\end{center}
\caption{Top panel: Temperature of the IGM around point sources with luminosities of $10^{54}~$ionizing photons per second and with spectral indexes of $\alpha = -0.5,~ -1.5$ and $-3$.  Each case is shown $10~$Myr and $100~$Myr after its source had turned on at $z=8$ in a fully neutral, homogeneous IGM.   Bottom Panel:  The same but where the emission is set to zero above $4~$Ry so that \HeII\ cannot be reionized.  In both panels, the vertical bars mark the location of the ionization front and are labeled by the velocity of the front.  As the front's velocity slows, the amount of cooling increases and the difference in the temperature between the different spectral models decreases.  The dotted curve in the top panel is the same as the $\alpha = -3$ case after $100$~Myr except there is a second, harder component with $\alpha=-0.5$ and comprising $10\%$ of the ionizing photons.  A luminosity of $10^{54}~$ionizing~photons~s$^{-1}$ is a few order of magnitudes less than that of the brightest $z\sim 6$ quasars and could represent the emission from a faint AGN or from an accumulation of galaxies (see text).
\label{fig:Thist}}
\end{figure}

We ran 1D radiative transfer calculations around a single point source in an initially neutral, homogeneous medium.  We assume spherical symmetry, a luminosity of $10^{54}$~ionizing photons per second, and that the source turned on at $z=8$.  We ignored adiabatic and Compton cooling as this prevents cooling well behind the front, aiding interpretation.  We will argue that a luminosity of $10^{54}$~photons s$^{-1}$ results in front speeds after $100$~Myr with similar velocities to those expected during cosmological reionization.  However, our results can essentially be rescaled to other source luminosities by taking the temperature at the radius with the same incident flux, $L/r^2$.  

Figure \ref{fig:Thist} shows the temperature profile from this calculation for sources with spectral indexes of $\alpha = -3, -1.5$ and $-0.5$.  The bottom panel is the same as the top panel except that a spectral cutoff is assumed above $4~$Ry, such that the helium cannot become doubly ionized.  Such a cutoff is meant to represent the case in which radiation from the starlight reionized the Universe.  Commonly, this case is assumed to have a very soft spectrum with $-\alpha = 3-4$, but recent spectral models are consistent with harder spectra (e.g., see discussion in \citealt{faucher09}),  and absorption within the host-galaxy should also harden the emission that reaches the IGM.  

In both panels in Figure \ref{fig:Thist}, the thin dashed curves show the temperature after $10~$Myr, when the ionization front is moving at $78~$cMpc~Gyr$^{-1}$, and the thick solid curves are the temperature after $100~$Myr, when the front is moving at $17~$cMpc~Gyr$^{-1}$.  The vertical bars mark the positions of the front (the \HII\ region radius).  The temperature immediately behind the ionization front is typically larger at earlier times, when the front was moving faster.  This temperature difference owes to the importance of collisional cooling at the ionization front -- the longer the front overlaps with a single gas element the more that element can cool (as the cooling rate peaks strongly at a hydrogen fraction of $0.5$).  Large front velocities enable reionization to heat the IGM to $>30,000~$K near bright sources (as can be seen in Fig. \ref{fig:Thist}; smaller radii have larger temperatures).

What speeds are anticipated for cosmological ionization fronts?
It is thought that the Universe was ionized by small galaxies with a number density of $n _g\sim 1~\cMpc^{-3}$, each producing on the order of $10^{51}$~ionizing photons per second (e.g., \citealt{mcquinn07}).  The front speed for an accumulation of galaxies with $n_g = 1~\cMpc^{-3}$ in a $\sim 5~\cMpc$ \HII\ region will be similar to that of the $100~\Myr$ curves in Figure \ref{fig:Thist}.  An alternative estimate for the ionization front speeds is given by cosmological simulations of reionization.  These simulations show that reionization occurs over $f \sim 0.5$ of a Hubble time and with characteristic bubble sizes of $R_b \sim 10~ \cMpc$, implying front speeds of $f^{-1} H^{-1} R_{b} \sim 20~$cMpc~Gyr$^{-1}$.  In detail, these simulations find that the fronts move slowly at first around individual sources but accelerate as more sources form and larger bubbles grow, having front speeds closer to $80~$cMpc~Gyr$^{-1}$ by the end \citep{iliev06, mcquinn07, trac07}.  Such acceleration would result in the regions that were ionized last being the warmest, especially for the harder spectra that we consider.  However, the ionization front speeds at the end of reionization in these simulations are almost certainly too fast, as these simulations are not capturing Lyman-limit systems, which absorb ionizing photons and slow the growth of \HII\ bubbles \citep{furlanetto05, choudhury09}.  In addition, the emissivity prescribed in these simulations tends to overshoot the observed ionizing emissivity at $z=6$ \citep{bolton07}, which also implies that reionization is happening too quickly in the simulations.  The longer the duration for reionization, the smaller the range of post-front temperatures. 


Thus, if the front speeds are of the order of $\sim 20~$cMpc~Gyr$^{-1}$, as we have argued, then the heating between the different models considered in Figure \ref{fig:Thist} is not substantially different, ranging between $17-25\times10^3~$K for the wide range of spectra that we considered.  Factors of $4$ faster speeds result in $2-3\times10^4~$K (thin dashed curves in the top panel of Fig. \ref{fig:Thist}), whereas slower speeds result in a slightly narrower range.  Note that the temperatures for the $\alpha=-1.5$ case can be larger than the harder $\alpha = -0.5$ case.   This difference owes to the larger amount of energy that was lost in secondary ionizations/excitations in the $\alpha = -0.5$ case, and it also owes to when the helium was reionized.  If helium is doubly ionized ahead of the hydrogen ionization front (as for $\alpha = -0.5$), then the imparted energy is more significantly radiated away at the time the hydrogen ionization front passes by collisional cooling \citep{miralda94}.  

Note that the case with reionization by sources with $\alpha = -0.5$ is ruled out by the SXB (\S \ref{sec:SXB}).  However, the scenario where $\approx10\%$ of the ionizations occurred via such a hard background is still allowed.  The dotted curve in the top panel is this case after $100$~Myr where $90\%$ of the photons are emitted with $\alpha = -3$ and $10\%$ with $\alpha = -0.5$.  The temperature of the IGM is only slightly increased in this case over reionization by a source with $\alpha = -3$.

For reionization by stellar radiation -- the most probable scenario --, it is unlikely that much $>4~$Ry radiation would have been produced (and, if it were, that it would escape into the IGM).  However, as mentioned above, there is significant uncertainty in the spectrum of the radiation that ultimately escapes a galaxy at $1-4~$Ry.  The bottom panel shows the temperatures in the case where the emergent spectrum has $\alpha = -0.5, -1.5$, and $-3$ at $1-4~$Ry.  This panel demonstrates that, despite the uncertainty in the emergent spectrum for stellar sources, there is little uncertainty in the temperature to which the IGM would have been heated in this scenario.

Our temperature profile calculations have been done in the limit of a homogeneous universe.  We argue in what follows that density inhomogeneities should not significantly alter our results.  In the limit of no recombinations (relevant for most inhomogeneities), the time an ionization front requires to pass a given point does not depend on these inhomogeneities.  While in this limit the front-crossing time is the same, the cooling rate per particle is proportional to density so that overdense regions would have cooled to somewhat lower temperatures as the front crossed.  In high enough density systems, recombinations can become important (which happens in overdensities of  $\sim 100~ \Gamma_{\rm HI}^{2/3}$, in  ``Lyman-limit systems''; \citealt{schaye01, mcquinn11}).  The continuum absorption of these systems can affect the post-front temperature, as they will slow the front (allowing for more cooling at the front) and harden the incident spectrum (so that more heat is injected per ionization).\footnote{In all modern models for reionization many sources resided in an \HII\ region during the bulk of reionization.  In this picture, most sources' radiation incident on a gas parcel at the ionization front will miss the Lyman-limit absorber that obscures a given sightline, lessening the impact of such radiative transfer effects.} Current thinking is that the number of ionizing photons required to ionized the universe is not much greater than one per hydrogen atom \citep{pawlik09, mcquinn11}.  This thinking is consistent with the finding that the Universe was not producing many more than one ionizing photon per Hubble time per baryon at $z=6$ \citep{miralda03, bolton07, mcquinn11}.  Thus, a large fraction of the emitting ionizing photons were unlikely to be absorbed in dense systems and change the character of the radiation field that reionized the IGM.  However, even in a universe where there were a substantial number of absorptions in Lyman-limit systems that hardened the incident spectrum, our calculations for different spectral hardnesses show that the range of possible post-front temperatures is still small.  This is particularly true for the (most plausible) case where reionization owes to $<4~$Ry photons.

Recent analyses have constrained the temperature that the IGM was heated to by hydrogen reionization.  The measurements of \citet{becker11} constrain the IGM temperature to $\approx 5-9 \times 10^3~$K at $z\approx 4.5$, and they are consistent with reionization heating the IGM to $20,000~$K at $z>8$.   In addition, \citet{bolton11} measured the temperature in the proximity regions of seven quasars at $z\approx6$, finding $T= 12-20 \times 10^4~$K at 95\%~C.L.   \citet{bolton11} concluded that this temperature indicates that \HeII\ was ionized by the quasar and that the hydrogen was reionized at $z>8$ (see also \citealt{raskutti12}, who accounted for more complex reionization histories).  Since we predict a small range in post-front temperatures with $T \approx 20,000$K, these constraints are also consistent with all the models considered here -- at least for the front speeds we argued are most applicable and as long as much of reionization occurs at $z>8$, the currently favored picture \citep{larson11}.

\section{Conclusions}

Several studies have suggested that the production of soft X-ray photons was more efficient per unit star formation rate at $z\sim10$ than it is at $z\sim 0$ and that these photons contributed significantly to the reionization of the Universe \citep{oh01,madau04, mirabel11, johnson11}.  In addition, reionization by quasars may not yet be ruled out.  While it has often been reasoned that quasars could not have reionized the Universe, this rationale has traditionally relied on observations of the decreasing number density of bright, $L>L_*$ quasars at $z>3$ \citep{madau99, faucher08b}.  This paper presented a census of constraints on high-redshift soft X-ray photon production.

We discussed the constraints from (1) the unresolved soft X-ray background (SXB), (2) intergalactic metal line observations, (3) the late reionization of helium, and (4) the temperature of the intergalactic medium.  We found that the unresolved SXB places interesting bounds on the contribution to reionization of sources with a spectral index in intensity per unit frequency, $\alpha$, greater than $ -1$.  This bound limits the total number of ionizing photons radiated per hydrogen atom to be $<1$ from several potential sources: inverse Compton scattering off supernova-accelerated electrons, HMXBs (under empirically motivated spectral models), and dark matter annihilations.  It limits the contribution to be $\ll 1$ in the latter two cases.  All of these sources have been proposed as important contributors to hydrogen reionization.  However, we showed that the SXB is consistent with quasars reionizing the Universe, correcting a common misperception.  In addition, we showed that if high-redshift galactic X-ray emissions account for the unresolved SXB intensity, significant evolution in hard X-ray luminosity--SFR relationship from that observed at low redshifts is required (unless there was more than an order of magnitude additional star formation at $z\sim 8$ than has been observed).

We showed that $z\approx 2.5$ intergalactic metal line observations (which constrain the coeval EUV background) are consistent with quasar-like sources with $\alpha \approx -1.7$ dominating the extragalactic background.  Metal absorption line observations do not allow a harder component than quasars to source more than half of the $\sim 200~$eV background at $z= 2.5$.  In addition, if a source population with $\alpha = -1$ contributed even just $10\%$ of the $1~$Ry emissivity at $z=2.5$, it would overshoot measured bounds on the unresolved SXB. 

We found that the contribution of faint AGN to high-redshift ionizing backgrounds is most constrained by our knowledge of the reionization history of hydrogen and helium (but recent luminosity function measurements also exclude much of the parameter space for AGN reionization; \citealt{willott10}).  We showed that quasars cannot finish ionizing the helium until $z=2.7$, while also ionizing the \HI\ at $z>6$, unless theoretical estimates for the clumpiness of intergalactic gas are low and the EUV emissivity of quasars/AGN stays constant or increases with increasing redshift.

Lastly, we investigated models for the heating of the IGM owing to reionization with a $1$D radiative transfer code.  Our calculations account for the finite speed of light as well as secondary heating and excitation processes, effects that had previously either been ignored or treated in an approximate manner, but are in fact important to include when soft X-ray photons contribute to reionization.  We argued that the bulk of the intergalactic gas would have been heated at reionization to temperatures greater than $17,000~$K and less than $25,000~$K for reionization by a wide range of spectral models (with an even narrower range if the sources cannot doubly ionize the helium).  The narrowness of this range owes to the efficiency of cooling at the ionization front for plausible front speeds.  The higher temperatures in this range did not necessarily imply that the gas was ionized by the hardest spectra, as our intermediate $\alpha=-1.5$ case resulted in the highest post-front temperatures.  In addition, the speed of the ionization front was as important for determining the post-front temperature as the emitted spectrum.   Therefore, it would be difficult to distinguish models for the sources of reionization based on measurements of the intergalactic thermal history.  Instead, the timing and duration of reionization is most important for establishing the average temperature of the post-reionization IGM.\\

We are grateful to Rosanne Di Stefano, Ryan O'Leary, Joop Schaye, and Jennifer Yeh for useful discussions.  We thank the Aspen Center for Physics (NSF grant \#1066293), where this work was initiated.  MM is supported by the National Aeronautics and Space
Administration through Einstein Postdoctoral Fellowship Award Number
PF9-00065 issued by the Chandra X-ray Observatory Center, which is
operated by the Smithsonian Astrophysical Observatory for and on
behalf of the National Aeronautics Space Administration under contract
NAS8-03060.

\bibliographystyle{apj}
\bibliography{Xrays}

\begin{thebibliography}{}

\bibitem[\protect\citeauthoryear{{Agafonova} et~al.}{{Agafonova}
  et~al.}{2007}]{agafonova07}
{Agafonova}, I.~I., {Levshakov}, S.~A., {Reimers}, D., {Fechner}, C., {Tytler},
  D., {Simcoe}, R.~A.,  \& {Songaila}, A. 2007, \aap, 461, 893

\bibitem[\protect\citeauthoryear{{Allen} et~al.}{{Allen}
  et~al.}{2008}]{allen08}
{Allen}, M.~G., {Groves}, B.~A., {Dopita}, M.~A., {Sutherland}, R.~S.,  \&
  {Kewley}, L.~J. 2008, \apjs, 178, 20

\bibitem[\protect\citeauthoryear{{Becker} et~al.}{{Becker}
  et~al.}{2011}]{becker11}
{Becker}, G.~D., {Bolton}, J.~S., {Haehnelt}, M.~G.,  \& {Sargent}, W.~L.~W.
  2011, \mnras, 410, 1096

\bibitem[\protect\citeauthoryear{{Belikov} \& {Hooper}}{{Belikov} \&
  {Hooper}}{2009}]{belikov09}
{Belikov}, A.~V.,  \& {Hooper}, D. 2009, \prd, 80, 035007

\bibitem[\protect\citeauthoryear{{Bolton} et~al.}{{Bolton}
  et~al.}{2011}]{bolton11}
{Bolton}, J.~S., {Becker}, G.~D., {Raskutti}, S., {Wyithe}, J.~S.~B.,
  {Haehnelt}, M.~G.,  \& {Sargent}, W.~L.~W. 2011, ArXiv:1110.0539

\bibitem[\protect\citeauthoryear{{Bolton} \& {Haehnelt}}{{Bolton} \&
  {Haehnelt}}{2007a}]{bolton07b}
{Bolton}, J.~S.,  \& {Haehnelt}, M.~G. 2007a, \mnras, 374, 493

\bibitem[\protect\citeauthoryear{{Bolton} \& {Haehnelt}}{{Bolton} \&
  {Haehnelt}}{2007b}]{bolton07}
{Bolton}, J.~S.,  \& {Haehnelt}, M.~G. 2007b, \mnras, 382, 325

\bibitem[\protect\citeauthoryear{{Bolton} et~al.}{{Bolton}
  et~al.}{2005}]{bolton05}
{Bolton}, J.~S., {Haehnelt}, M.~G., {Viel}, M.,  \& {Springel}, V. 2005,
  \mnras, 357, 1178

\bibitem[\protect\citeauthoryear{{Bouwens} et~al.}{{Bouwens}
  et~al.}{2011}]{bouwens11b}
{Bouwens}, R.~J., et~al. 2011, \nat, 469, 504

\bibitem[\protect\citeauthoryear{{Cantalupo} \& {Porciani}}{{Cantalupo} \&
  {Porciani}}{2011}]{cantalupo11}
{Cantalupo}, S.,  \& {Porciani}, C. 2011, \mnras, 411, 1678

\bibitem[\protect\citeauthoryear{{Cen} \& {Ostriker}}{{Cen} \&
  {Ostriker}}{1999}]{cen99}
{Cen}, R.,  \& {Ostriker}, J.~P. 1999, \apj, 514, 1

\bibitem[\protect\citeauthoryear{{Chen}, {Prochaska}, \& {Gnedin}}{{Chen}
  et~al.}{2007}]{chen07}
{Chen}, H.-W., {Prochaska}, J.~X.,  \& {Gnedin}, N.~Y. 2007, \apjl, 667, L125

\bibitem[\protect\citeauthoryear{{Choudhury}, {Haehnelt}, \&
  {Regan}}{{Choudhury} et~al.}{2009}]{choudhury09}
{Choudhury}, T.~R., {Haehnelt}, M.~G.,  \& {Regan}, J. 2009, \mnras, 394, 960

\bibitem[\protect\citeauthoryear{{Cowie}, {Barger}, \& {Hasinger}}{{Cowie}
  et~al.}{2012}]{cowie11}
{Cowie}, L.~L., {Barger}, A.~J.,  \& {Hasinger}, G. 2012, \apj, 748, 50

\bibitem[\protect\citeauthoryear{{Crowther} et~al.}{{Crowther}
  et~al.}{2010}]{crowther10}
{Crowther}, P.~A., {Barnard}, R., {Carpano}, S., {Clark}, J.~S., {Dhillon},
  V.~S.,  \& {Pollock}, A.~M.~T. 2010, \mnras, 403, L41

\bibitem[\protect\citeauthoryear{{Dijkstra} et~al.}{{Dijkstra}
  et~al.}{2012}]{dijkstra11}
{Dijkstra}, M., {Gilfanov}, M., {Loeb}, A.,  \& {Sunyaev}, R. 2012, \mnras,
  421, 213

\bibitem[\protect\citeauthoryear{{Dijkstra}, {Haiman}, \& {Loeb}}{{Dijkstra}
  et~al.}{2004}]{dijkstra04}
{Dijkstra}, M., {Haiman}, Z.,  \& {Loeb}, A. 2004, \apj, 613, 646

\bibitem[\protect\citeauthoryear{{Dijkstra} \& {Wyithe}}{{Dijkstra} \&
  {Wyithe}}{2006}]{dijkstra06}
{Dijkstra}, M.,  \& {Wyithe}, J.~S.~B. 2006, \mnras, 372, 1575

\bibitem[\protect\citeauthoryear{{Dopita} et~al.}{{Dopita}
  et~al.}{2011}]{dopita11}
{Dopita}, M.~A., {Krauss}, L.~M., {Sutherland}, R.~S., {Kobayashi}, C.,  \&
  {Lineweaver}, C.~H. 2011, \apss, 358

\bibitem[\protect\citeauthoryear{{Fan} et~al.}{{Fan} et~al.}{2001}]{fan01}
{Fan}, X.,  et~al. 2001, \aj, 122, 2833

\bibitem[\protect\citeauthoryear{{Fan} et~al.}{{Fan} et~al.}{2006}]{fan06}
{Fan}, X., et~al. 2006, \aj, 132, 117

\bibitem[\protect\citeauthoryear{{Faucher-Gigu{\`e}re}
  et~al.}{{Faucher-Gigu{\`e}re} et~al.}{2008a}]{faucher08b}
{Faucher-Gigu{\`e}re}, C., {Lidz}, A., {Hernquist}, L.,  \& {Zaldarriaga}, M.
  2008a, \apj

\bibitem[\protect\citeauthoryear{{Faucher-Gigu{\`e}re}
  et~al.}{{Faucher-Gigu{\`e}re} et~al.}{2009}]{faucher09}
{Faucher-Gigu{\`e}re}, C., {Lidz}, A., {Zaldarriaga}, M.,  \& {Hernquist}, L.
  2009, \apj, 703, 1416

\bibitem[\protect\citeauthoryear{{Faucher-Gigu{\`e}re}
  et~al.}{{Faucher-Gigu{\`e}re} et~al.}{2008b}]{faucher08}
{Faucher-Gigu{\`e}re}, C., {Prochaska}, J.~X., {Lidz}, A., {Hernquist}, L.,  \&
  {Zaldarriaga}, M. 2008b, \apj

\bibitem[\protect\citeauthoryear{{Fechner}}{{Fechner}}{2011}]{fechner11}
{Fechner}, C. 2011, \aap, 532, A62

\bibitem[\protect\citeauthoryear{{Finlator}, {Dav{\'e}}, \&
  {{\"O}zel}}{{Finlator} et~al.}{2011}]{finlator11b}
{Finlator}, K., {Dav{\'e}}, R.,  \& {{\"O}zel}, F. 2011, \apj

\bibitem[\protect\citeauthoryear{{Finlator}, {Oppenheimer}, \&
  {Dav{\'e}}}{{Finlator} et~al.}{2011}]{finlator11}
{Finlator}, K., {Oppenheimer}, B.~D.,  \& {Dav{\'e}}, R. 2011, \mnras

\bibitem[\protect\citeauthoryear{{Furlanetto} \& {Dixon}}{{Furlanetto} \&
  {Dixon}}{2010}]{furlanettodixon}
{Furlanetto}, S.~R.,  \& {Dixon}, K.~L. 2010, \apj, 714, 355

\bibitem[\protect\citeauthoryear{{Furlanetto} \& {Oh}}{{Furlanetto} \&
  {Oh}}{2005}]{furlanetto05}
{Furlanetto}, S.~R.,  \& {Oh}, S.~P. 2005, \mnras, 363, 1031

\bibitem[\protect\citeauthoryear{{Furlanetto} \& {Oh}}{{Furlanetto} \&
  {Oh}}{2009}]{furlanetto09}
{Furlanetto}, S.~R.,  \& {Oh}, S.~P. 2009, \apj, 701, 94

\bibitem[\protect\citeauthoryear{{Furlanetto}, {Oh}, \& {Briggs}}{{Furlanetto}
  et~al.}{2006}]{furlanettoohbriggs}
{Furlanetto}, S.~R., {Oh}, S.~P.,  \& {Briggs}, F.~H. 2006, \physrep, 433, 181

\bibitem[\protect\citeauthoryear{{Furlanetto} \& {Stoever}}{{Furlanetto} \&
  {Stoever}}{2010}]{furlanetto10}
{Furlanetto}, S.~R.,  \& {Stoever}, S.~J. 2010, \mnras, 404, 1869

\bibitem[\protect\citeauthoryear{{Furlanetto}, {Zaldarriaga}, \&
  {Hernquist}}{{Furlanetto} et~al.}{2004}]{furlanetto04}
{Furlanetto}, S.~R., {Zaldarriaga}, M.,  \& {Hernquist}, L. 2004, \apj, 613, 1

\bibitem[\protect\citeauthoryear{{Glikman} et~al.}{{Glikman}
  et~al.}{2011}]{glikman11}
{Glikman}, E., {Djorgovski}, S.~G., {Stern}, D., {Dey}, A., {Jannuzi}, B.~T.,
  \& {Lee}, K.-S. 2011, \apjl, 728, L26

\bibitem[\protect\citeauthoryear{{Gnedin}}{{Gnedin}}{2008}]{gnedin08}
{Gnedin}, N.~Y. 2008, \apjl, 673, L1

\bibitem[\protect\citeauthoryear{{Gnedin}, {Kravtsov}, \& {Chen}}{{Gnedin}
  et~al.}{2008}]{gnedin08b}
{Gnedin}, N.~Y., {Kravtsov}, A.~V.,  \& {Chen}, H.-W. 2008, \apj, 672, 765

\bibitem[\protect\citeauthoryear{{Haardt} \& {Madau}}{{Haardt} \&
  {Madau}}{1996}]{haardt96}
{Haardt}, F.,  \& {Madau}, P. 1996, \apj, 461, 20

\bibitem[\protect\citeauthoryear{{Haardt} \& {Madau}}{{Haardt} \&
  {Madau}}{2012}]{haardt11}
{Haardt}, F.,  \& {Madau}, P. 2012, \apj, 746, 125

\bibitem[\protect\citeauthoryear{{Haiman}}{{Haiman}}{2011}]{haimanopinion}
{Haiman}, Z. 2011, \nat, 472, 47

\bibitem[\protect\citeauthoryear{{Hickox} \& {Markevitch}}{{Hickox} \&
  {Markevitch}}{2007}]{hickox07}
{Hickox}, R.~C.,  \& {Markevitch}, M. 2007, \apjl, 661, L117

\bibitem[\protect\citeauthoryear{{Hopkins}, {Richards}, \&
  {Hernquist}}{{Hopkins} et~al.}{2007}]{hopkins07}
{Hopkins}, P.~F., {Richards}, G.~T.,  \& {Hernquist}, L. 2007, \apj, 654, 731

\bibitem[\protect\citeauthoryear{{Hui} \& {Gnedin}}{{Hui} \&
  {Gnedin}}{1997}]{hui97}
{Hui}, L.,  \& {Gnedin}, N.~Y. 1997, \mnras, 292, 27

\bibitem[\protect\citeauthoryear{{Hui} \& {Haiman}}{{Hui} \&
  {Haiman}}{2003}]{hui03}
{Hui}, L.,  \& {Haiman}, Z. 2003, \apj, 596, 9

\bibitem[\protect\citeauthoryear{{Iliev} et~al.}{{Iliev}
  et~al.}{2006}]{iliev06}
{Iliev}, I.~T., {Mellema}, G., {Pen}, U., {Merz}, H., {Shapiro}, P.~R.,  \&
  {Alvarez}, M.~A. 2006, \mnras, 369, 1625

\bibitem[\protect\citeauthoryear{{Johnson} \& {Khochfar}}{{Johnson} \&
  {Khochfar}}{2011}]{johnson11}
{Johnson}, J.~L.,  \& {Khochfar}, S. 2011, \apj, 743, 126

\bibitem[\protect\citeauthoryear{{Kaaret}, {Schmitt}, \& {Gorski}}{{Kaaret}
  et~al.}{2011}]{kaaret11}
{Kaaret}, P., {Schmitt}, J.,  \& {Gorski}, M. 2011, \apj, 741, 10

\bibitem[\protect\citeauthoryear{{Kobayashi} et~al.}{{Kobayashi}
  et~al.}{2004}]{kobayashi04}
{Kobayashi}, T., {Komori}, Y., {Yoshida}, K.,  \& {Nishimura}, J. 2004, \apj,
  601, 340

\bibitem[\protect\citeauthoryear{{Kuhlen} \& {Faucher-Gigu{\`e}re}}{{Kuhlen} \&
  {Faucher-Gigu{\`e}re}}{2012}]{kuhlen12}
{Kuhlen}, M.,  \& {Faucher-Gigu{\`e}re}, C.-A. 2012, \mnras, 423, 862

\bibitem[\protect\citeauthoryear{{Larson} et~al.}{{Larson}
  et~al.}{2011}]{larson11}
{Larson}, D., et~al. 2011, \apjs, 192, 16

\bibitem[\protect\citeauthoryear{{Lidz} et~al.}{{Lidz} et~al.}{2007}]{lidz07}
{Lidz}, A., {McQuinn}, M., {Zaldarriaga}, M., {Hernquist}, L.,  \& {Dutta}, S.
  2007, \apj, 670, 39

\bibitem[\protect\citeauthoryear{{Lutovinov} et~al.}{{Lutovinov}
  et~al.}{2005}]{lutovinov05}
{Lutovinov}, A., {Revnivtsev}, M., {Gilfanov}, M., {Shtykovskiy}, P., {Molkov},
  S.,  \& {Sunyaev}, R. 2005, \aap, 444, 821

\bibitem[\protect\citeauthoryear{{Madau}, {Haardt}, \& {Rees}}{{Madau}
  et~al.}{1999}]{madau99}
{Madau}, P., {Haardt}, F.,  \& {Rees}, M.~J. 1999, \apj, 514, 648

\bibitem[\protect\citeauthoryear{{Madau} et~al.}{{Madau}
  et~al.}{2004}]{madau04}
{Madau}, P., {Rees}, M.~J., {Volonteri}, M., {Haardt}, F.,  \& {Oh}, S.~P.
  2004, \apj, 604, 484

\bibitem[\protect\citeauthoryear{{McQuinn}}{{McQuinn}}{2009}]{mcquinnGP}
{McQuinn}, M. 2009, \apjl, 704, L89

\bibitem[\protect\citeauthoryear{{McQuinn} et~al.}{{McQuinn}
  et~al.}{2007}]{mcquinn07}
{McQuinn}, M., {Lidz}, A., {Zahn}, O., {Dutta}, S., {Hernquist}, L.,  \&
  {Zaldarriaga}, M. 2007, \mnras, 377, 1043

\bibitem[\protect\citeauthoryear{{McQuinn} et~al.}{{McQuinn}
  et~al.}{2009}]{mcquinn09}
{McQuinn}, M., {Lidz}, A., {Zaldarriaga}, M., {Hernquist}, L., {Hopkins},
  P.~F., {Dutta}, S.,  \& {Faucher-Gigu{\`e}re}, C. 2009, \apj, 694, 842

\bibitem[\protect\citeauthoryear{{McQuinn}, {Oh}, \&
  {Faucher-Gigu{\`e}re}}{{McQuinn} et~al.}{2011}]{mcquinn11}
{McQuinn}, M., {Oh}, S.~P.,  \& {Faucher-Gigu{\`e}re}, C.-A. 2011, \apj, 743,
  82

\bibitem[\protect\citeauthoryear{{McQuinn} \& {O'Leary}}{{McQuinn} \&
  {O'Leary}}{2012}]{mcquinn12}
{McQuinn}, M.,  \& {O'Leary}, R.~M. 2012, ArXiv:1204.1345

\bibitem[\protect\citeauthoryear{{McQuinn} \& {Switzer}}{{McQuinn} \&
  {Switzer}}{2010}]{mcquinnHeI}
{McQuinn}, M.,  \& {Switzer}, E.~R. 2010, \mnras, 408, 1945

\bibitem[\protect\citeauthoryear{{McQuinn} \& {Zaldarriaga}}{{McQuinn} \&
  {Zaldarriaga}}{2011}]{mcquinnHAZE}
{McQuinn}, M.,  \& {Zaldarriaga}, M. 2011, \mnras, 414, 3577

\bibitem[\protect\citeauthoryear{{Meiksin}}{{Meiksin}}{2009}]{meiksin09}
{Meiksin}, A.~A. 2009, Reviews of Modern Physics, 81, 1405

\bibitem[\protect\citeauthoryear{{Mesinger}}{{Mesinger}}{2010}]{mesinger10}
{Mesinger}, A. 2010, \mnras, 407, 1328

\bibitem[\protect\citeauthoryear{{Mesinger}, {McQuinn}, \&
  {Spergel}}{{Mesinger} et~al.}{2012}]{mesinger11}
{Mesinger}, A., {McQuinn}, M.,  \& {Spergel}, D.~N. 2012, \mnras, 422, 1403

\bibitem[\protect\citeauthoryear{{Mineo}, {Gilfanov}, \& {Sunyaev}}{{Mineo}
  et~al.}{2012}]{mineo11}
{Mineo}, S., {Gilfanov}, M.,  \& {Sunyaev}, R. 2012, \mnras, 419, 2095

\bibitem[\protect\citeauthoryear{{Mirabel} et~al.}{{Mirabel}
  et~al.}{2011}]{mirabel11}
{Mirabel}, I.~F., {Dijkstra}, M., {Laurent}, P., {Loeb}, A.,  \& {Pritchard},
  J.~R. 2011, \aap, 528, A149

\bibitem[\protect\citeauthoryear{{Miralda-Escud{\'e}}}{{Miralda-Escud{\'e}}}{2003}]{miralda03}
{Miralda-Escud{\'e}}, J. 2003, \apj, 597, 66

\bibitem[\protect\citeauthoryear{{Miralda-Escud{\'e}} \&
  {Rees}}{{Miralda-Escud{\'e}} \& {Rees}}{1994}]{miralda94}
{Miralda-Escud{\'e}}, J.,  \& {Rees}, M.~J. 1994, \mnras, 266, 343

\bibitem[\protect\citeauthoryear{{Moretti} et~al.}{{Moretti}
  et~al.}{2003}]{moretti03}
{Moretti}, A., {Campana}, S., {Lazzati}, D.,  \& {Tagliaferri}, G. 2003, \apj,
  588, 696

\bibitem[\protect\citeauthoryear{{Mortlock} et~al.}{{Mortlock}
  et~al.}{2011}]{mortlock11}
{Mortlock}, D.~J., et~al. 2011, \nat, 474, 616

\bibitem[\protect\citeauthoryear{{Nandra} et~al.}{{Nandra}
  et~al.}{2002}]{nandra02}
{Nandra}, K., {Mushotzky}, R.~F., {Arnaud}, K., {Steidel}, C.~C., {Adelberger},
  K.~L., {Gardner}, J.~P., {Teplitz}, H.~I.,  \& {Windhorst}, R.~A. 2002, \apj,
  576, 625

\bibitem[\protect\citeauthoryear{{Nestor} et~al.}{{Nestor}
  et~al.}{2011}]{nestor11}
{Nestor}, D.~B., {Shapley}, A.~E., {Steidel}, C.~C.,  \& {Siana}, B. 2011,
  \apj, 736, 18

\bibitem[\protect\citeauthoryear{{Oh}}{{Oh}}{2001}]{oh01}
{Oh}, S.~P. 2001, \apj, 553, 499

\bibitem[\protect\citeauthoryear{{Pawlik}, {Schaye}, \& {van
  Scherpenzeel}}{{Pawlik} et~al.}{2009}]{pawlik09}
{Pawlik}, A.~H., {Schaye}, J.,  \& {van Scherpenzeel}, E. 2009, \mnras, 394,
  1812

\bibitem[\protect\citeauthoryear{{Peebles}}{{Peebles}}{1993}]{peebles93}
 1993, {Principles of Physical Cosmology}, ed. {Peebles, P.~J.~E.}

\bibitem[\protect\citeauthoryear{{Power} et~al.}{{Power}
  et~al.}{2009}]{power09}
{Power}, C., {Wynn}, G.~A., {Combet}, C.,  \& {Wilkinson}, M.~I. 2009, \mnras,
  395, 1146

\bibitem[\protect\citeauthoryear{{Prochaska}, {O'Meara}, \&
  {Worseck}}{{Prochaska} et~al.}{2010}]{prochaska10}
{Prochaska}, J.~X., {O'Meara}, J.~M.,  \& {Worseck}, G. 2010, \apj, 718, 392

\bibitem[\protect\citeauthoryear{{Prochaska}, {Worseck}, \&
  {O'Meara}}{{Prochaska} et~al.}{2009}]{prochaska09}
{Prochaska}, J.~X., {Worseck}, G.,  \& {O'Meara}, J.~M. 2009, \apjl, 705, L113

\bibitem[\protect\citeauthoryear{{Raskutti} et~al.}{{Raskutti}
  et~al.}{2012}]{raskutti12}
{Raskutti}, S., {Bolton}, J.~S., {Wyithe}, J.~S.~B.,  \& {Becker}, G.~D. 2012,
  ArXiv:1201.5138

\bibitem[\protect\citeauthoryear{{Reddy} \& {Steidel}}{{Reddy} \&
  {Steidel}}{2004}]{reddy04}
{Reddy}, N.~A.,  \& {Steidel}, C.~C. 2004, \apjl, 603, L13

\bibitem[\protect\citeauthoryear{{Reichardt} et~al.}{{Reichardt}
  et~al.}{2012}]{reichardt11}
{Reichardt}, C.~L.,  et~al. 2012, \apj, 755, 70

\bibitem[\protect\citeauthoryear{{Rephaeli}, {Gruber}, \& {Persic}}{{Rephaeli}
  et~al.}{1995}]{rephaeli95}
{Rephaeli}, Y., {Gruber}, D.,  \& {Persic}, M. 1995, \aap, 300, 91

\bibitem[\protect\citeauthoryear{{Ricotti}, {Gnedin}, \& {Shull}}{{Ricotti}
  et~al.}{2002}]{ricotti02}
{Ricotti}, M., {Gnedin}, N.~Y.,  \& {Shull}, J.~M. 2002, \apj, 575, 49

\bibitem[\protect\citeauthoryear{{Salvaterra}, {Haardt}, \&
  {Volonteri}}{{Salvaterra} et~al.}{2007}]{salvaterra07}
{Salvaterra}, R., {Haardt}, F.,  \& {Volonteri}, M. 2007, \mnras, 374, 761

\bibitem[\protect\citeauthoryear{{Schaye}}{{Schaye}}{2001}]{schaye01}
{Schaye}, J. 2001, \apj, 559, 507

\bibitem[\protect\citeauthoryear{{Scott} et~al.}{{Scott}
  et~al.}{2004}]{scott04}
{Scott}, J.~E., {Kriss}, G.~A., {Brotherton}, M., {Green}, R.~F., {Hutchings},
  J., {Shull}, J.~M.,  \& {Zheng}, W. 2004, \apj, 615, 135

\bibitem[\protect\citeauthoryear{{Shankar} \& {Mathur}}{{Shankar} \&
  {Mathur}}{2007}]{shankar07}
{Shankar}, F.,  \& {Mathur}, S. 2007, \apj, 660, 1051

\bibitem[\protect\citeauthoryear{{Shull} \& {van Steenberg}}{{Shull} \& {van
  Steenberg}}{1985}]{shull85}
{Shull}, J.~M.,  \& {van Steenberg}, M.~E. 1985, \apj, 298, 268

\bibitem[\protect\citeauthoryear{{Shull} et~al.}{{Shull}
  et~al.}{2010}]{shull10}
{Shull}, M., {France}, K., {Danforth}, C., {Smith}, B.,  \& {Tumlinson}, J.
  2010, ArXiv:1008.2957

\bibitem[\protect\citeauthoryear{{Siana} et~al.}{{Siana}
  et~al.}{2008}]{siana08}
{Siana}, B., et~al. 2008, \apj, 675, 49

\bibitem[\protect\citeauthoryear{{Strong}, {Moskalenko}, \& {Reimer}}{{Strong}
  et~al.}{2004}]{strong04}
{Strong}, A.~W., {Moskalenko}, I.~V.,  \& {Reimer}, O. 2004, \apj, 613, 962

\bibitem[\protect\citeauthoryear{{Swartz} et~al.}{{Swartz}
  et~al.}{2004}]{swartz04}
{Swartz}, D.~A., {Ghosh}, K.~K., {Tennant}, A.~F.,  \& {Wu}, K. 2004, \apjs,
  154, 519

\bibitem[\protect\citeauthoryear{{Telfer} et~al.}{{Telfer}
  et~al.}{2002}]{telfer02}
{Telfer}, R.~C., {Zheng}, W., {Kriss}, G.~A.,  \& {Davidsen}, A.~F. 2002, \apj,
  565, 773

\bibitem[\protect\citeauthoryear{{Theuns} et~al.}{{Theuns}
  et~al.}{2002}]{theuns02}
{Theuns}, T., {Schaye}, J., {Zaroubi}, S., {Kim}, T.-S., {Tzanavaris}, P.,  \&
  {Carswell}, B. 2002, \apjl, 567, L103

\bibitem[\protect\citeauthoryear{{Thompson} et~al.}{{Thompson}
  et~al.}{2006}]{thompson06}
{Thompson}, T.~A., {Quataert}, E., {Waxman}, E., {Murray}, N.,  \& {Martin},
  C.~L. 2006, \apj, 645, 186

\bibitem[\protect\citeauthoryear{{Tittley} \& {Meiksin}}{{Tittley} \&
  {Meiksin}}{2007}]{tittley07}
{Tittley}, E.~R.,  \& {Meiksin}, A. 2007, \mnras, 380, 1369

\bibitem[\protect\citeauthoryear{{Tozzi} et~al.}{{Tozzi}
  et~al.}{2006}]{tozzi06}
{Tozzi}, P., et~al. 2006, \aap, 451, 457

\bibitem[\protect\citeauthoryear{{Trac} \& {Cen}}{{Trac} \&
  {Cen}}{2007}]{trac07}
{Trac}, H.,  \& {Cen}, R. 2007, \apj, 671, 1

\bibitem[\protect\citeauthoryear{{Trac}, {Cen}, \& {Loeb}}{{Trac}
  et~al.}{2008}]{trac08}
{Trac}, H., {Cen}, R.,  \& {Loeb}, A. 2008, \apjl, 689, L81

\bibitem[\protect\citeauthoryear{{Vanzella} et~al.}{{Vanzella}
  et~al.}{2010}]{vanzella10}
{Vanzella}, E., et~al. 2010, \apj, 725, 1011

\bibitem[\protect\citeauthoryear{{Venkatesan} \& {Benson}}{{Venkatesan} \&
  {Benson}}{2011}]{venkaesan11}
{Venkatesan}, A.,  \& {Benson}, A. 2011, \mnras, 417, 2264

\bibitem[\protect\citeauthoryear{{Venkatesan}, {Giroux}, \&
  {Shull}}{{Venkatesan} et~al.}{2001}]{venkatesan01}
{Venkatesan}, A., {Giroux}, M.~L.,  \& {Shull}, J.~M. 2001, \apj, 563, 1

\bibitem[\protect\citeauthoryear{{Volonteri} \& {Gnedin}}{{Volonteri} \&
  {Gnedin}}{2009}]{volonteri09}
{Volonteri}, M.,  \& {Gnedin}, N.~Y. 2009, \apj, 703, 2113

\bibitem[\protect\citeauthoryear{{Volonteri}, {Lodato}, \&
  {Natarajan}}{{Volonteri} et~al.}{2008}]{volonteri08}
{Volonteri}, M., {Lodato}, G.,  \& {Natarajan}, P. 2008, \mnras, 383, 1079

\bibitem[\protect\citeauthoryear{{Willott} et~al.}{{Willott}
  et~al.}{2010a}]{willott10b}
{Willott}, C.~J., et~al. 2010a, \aj

\bibitem[\protect\citeauthoryear{{Willott} et~al.}{{Willott}
  et~al.}{2010b}]{willott10}
{Willott}, C.~J., et~al. 2010b, \aj

\bibitem[\protect\citeauthoryear{{Worseck} et~al.}{{Worseck}
  et~al.}{2011}]{worseck11}
{Worseck}, G., et~al. 2011, \apjl, 733, L24

\bibitem[\protect\citeauthoryear{{Wyithe} \& {Loeb}}{{Wyithe} \&
  {Loeb}}{2003}]{wyithe03}
{Wyithe}, J.~S.~B.,  \& {Loeb}, A. 2003, \apj, 586, 693

\bibitem[\protect\citeauthoryear{{Wyithe}, {Mould}, \& {Loeb}}{{Wyithe}
  et~al.}{2011}]{wyithe11}
{Wyithe}, S., {Mould}, J.,  \& {Loeb}, A. 2011, ArXiv:1108.5809

\bibitem[\protect\citeauthoryear{{Yu} \& {Tremaine}}{{Yu} \&
  {Tremaine}}{2002}]{yu02}
{Yu}, Q.,  \& {Tremaine}, S. 2002, \mnras, 335, 965

\end{thebibliography}

\end{document}